\documentstyle[12pt]{article}
\begin{document}
\thispagestyle{empty}
\begin{flushright}\begin{tabular}{l}
hep-th/9605159\\
PAR-LPTHE-96-14 \\
ULB-TH-96/06\\
IMAFF-RCA-96-01
\end{tabular}\end{flushright}
\vspace{5mm}
\begin{center}
{\LARGE\bf The dynamical structure of higher  \\[5mm] 
dimensional Chern-Simons theory}\\[10mm]
{M\'aximo Ba\~nados$^a$, Luis J. Garay$^b$ and 
Marc Henneaux$^{a,c}$ } \\[5mm]
{\em $^{(a)}$ Centro de Estudios
Cient\'{\i}ficos de Santiago, Casilla 16443, Santiago, Chile \\
$^{(b)}$ Centro de F\'{\i}sica Miguel A. Catal\'an,
Instituto de Matem\'aticas y F\'{\i}sica Fundamental,
CSIC, Serrano 121, E--28006 Madrid, Spain  \\ 
$^{(c)}$ Laboratoire de Physique Th\'eorique et Hautes Energies, \\
Universit\'es Paris VI et Paris VII, Bte 126, 4, Place Jussieu,\\
75252 Paris Cedex 05, France }\footnote{Permanent
address:  Universit\'e
Libre de Bruxelles, Campus Plaine, C.P. 231, B-1050, Bruxelles,
Belgium}\\[5mm]
29 May 1996
\end{center}
\vspace{.5cm}
\begin{abstract}

Higher dimensional Chern-Simons theories, even though constructed along
the same topological pattern as in 2+1 dimensions, have been shown
recently to have generically a non-vanishing number of degrees of
freedom. In this paper, we carry out the complete Dirac Hamiltonian
analysis (separation of first and second class constraints and
calculation of the Dirac bracket) for a group $G\times U(1)$. We also
study the algebra of surface charges that arise in the presence of
boundaries and show that it is isomorphic to the WZW$_4$ discussed in
the literature. Some applications are then considered. It is shown, in
particular, that Chern-Simons gravity in dimensions greater than or
equal to five has a propagating torsion.

\end{abstract}

\vfill
\break

\renewcommand{\theequation}{\arabic{section}.\arabic{equation}}

\section{Introduction}
\setcounter{equation}{0}

In a previous paper \cite{BGH}, we have shown that pure
Chern-Simons theories in spacetime dimensions greater than or
equal to five possess local degrees of freedom in contrast to
the familiar three dimensional case. The only exception is the
Chern-Simons theory based on the one-dimensional group $U(1)$,
which is devoid of local degrees of freedom for any spacetime
dimension. However, whenever the gauge group is of dimension
greater than 1, the Chern-Simons action generically contains
propagating degrees of freedom.

One way to understand this somewhat unexpected result is to
observe that the equations of motion no longer imply that the
curvature vanishes for dimensions greater than or equal to five.
If there is only one gauge field, i.e., if the gauge group is
$G=U(1)$, one can always bring its curvature $F$ to some fixed
canonical form by using the diffeomorphism invariance (Darboux
theorem). Thus, even though $F\neq 0$, one may assume that it
has a fixed form and, therefore, the space of solutions of the
equations of motion modulo gauge transformations is reduced to a
single point. If, however, the gauge group is larger, there are
more curvatures.  One may bring one of them to a fixed canonical
form as for $U(1)$, but once this is done, there is not enough
invariance left to fix the other curvatures in a similar way.
Thus, the space of solutions is now bigger and, consequently,
there exist local degrees of freedom.

The number of local degrees of freedom was explicitly counted in
Ref. \cite{BGH} by using the Hamiltonian formalism. The phase
space of the theory was constructed and all the constraints were
exhibited. We also derived the number of second and first class
constraints. This provided the necessary information to count
the number of local degrees of freedom according to the formula,
\begin{equation} 
{\cal N} =\frac{1}{2} (P-2F-S),
\end{equation} 
where $P$ is the dimension of phase space, $F$ is the number of
first class constraints and $S$ is the number of second class
constraints (see, e.g., Ref. \cite{HT-book}).

Although the analysis of Ref. \cite{BGH} enables one conclude
rigorously  to the existence of local degrees of freedom, the
second class constraints were not explicitly separated from the
first class ones; only their number was given and the Dirac
bracket associated to the elimination of the second class
constraints was not computed either. As it is known, these steps
are quite important and must be carried out before trying to
quantize theory.

An interesting feature of higher-dimensional Chern-Simons
theories, also displayed in Ref. \cite{BGH}, is that these are
theories invariant under spacetime diffeomorphisms {\em for
which the generator ${\cal H}({\bf x})$ of timelike
diffeomorphisms is not independent from the other constraints}.
That is, the timelike diffeomorphisms can be expressed in terms
of the spacelike diffeomorphisms and of the internal gauge
transformations.  Now, it is well known that the
``superhamiltonian" constraint ${\cal H}({\bf x}) |\psi\rangle = 0$ is
usually the one that resists an exact treatment in the quantum
theory.  This appears quite strikingly in the loop
representation approach to quantum gravity \cite{Ashtekar}.  In
the case of higher-dimensional Chern-Simons theories, however,
the ``hard constraints" are absent, even though there are local
degrees of freedom.  Thus, any state that is invariant under
both the internal gauge symmetries and the spacelike
diffeomorphisms is automatically a solution of all the quantum
constraint equations.  Solving the ``kinematical" constraints
associated with spacelike diffeomorphism invariance and internal
gauge invariance, is, however, not entirely straightforward in
higher-dimensional Chern-Simons theories.  This is because the
components of the connection have non trivial Dirac brackets and
thus do not define simultaneously diagonalizable operators. It
makes the issue of computing these brackets even more pressing.
 
In this paper, we complete the canonical analysis of higher
dimensional Chern-Simons theory. It turns out that the
separation of first and second class constraints is technically
intricate for an arbitrary gauge group $G$. However, if one
considers a Chern-Simons action for the gauge group $G\times
U(1)$, the calculations become much simpler.  The only
requirement on the group $G$ is that it possesses a
non-degenerate bilinear invariant form. In order to avoid
uninteresting (and conceptually trivial) complications, we shall
therefore complete the Dirac bracket analysis only in that case,
as well as in the $U(1)$-case, which has its own peculiarities.

Three dimensional Chern-Simons theory is well known to induce a rich
dynamics at the boundary \cite{Moore-Sieberg}. One may therefore wonder
whether this is also the case in higher dimensional spacetimes. The
answer to this questions is affirmative. This problem has been already
considered in the literature \cite{Dunne,Nair}. Specifically, in Ref.
\cite{Nair}, the connection between a `conformal' field theory in four
dimensions and a K\"ahler five dimensional Chern-Simons theory was
established.  In this work, we explicitly exhibit the
symmetry algebra arising at the boundary for the full Chern-Simons
theory with no extra assumption other than the boundary conditions. For
definiteness, we consider Chern-Simons theory in five dimensions and
show that, if the gauge group is taken to be $G\times U(1)$, as above,
then the resulting symmetry algebra is just the WZW$_4$ algebra (based
on $G$) discussed in Ref. \cite{Nair}, with the curvature of the $U(1)$
factor appearing as a K\"ahler form. [For a recent work dealing with the
WZW$_4$ algebra see Ref. \cite{Moore-}.]
 
This paper is organized as follows. After a brief survey of our
conventions, we review in Sec. \ref{paper1} the results of Ref.
\cite{BGH}. We recall, in particular, the importance of the
so-called {\em generic condition} that was introduced there.
Then, we illustrate the generic condition in the physically
interesting context of Lovelock-Chern-Simons gravity (Sec.
\ref{Gravity}), as well as for some seven dimensional
Chern-Simons theories (Sec. \ref{D=7}). In Sec. \ref{N=1}, we
complete the Dirac analysis for the $U(1)$ theory, which has no
degrees of freedom. We show that in this theory one needs to
break general covariance in order to separate the first and
second class constraints. This is quite analogous to what
happens for the superparticle \cite{Bengtsson,Brink,Kall} and
the analogies are pointed out. We turn next (Sec. \ref{G*U(1)})
to the separation of first and second class constraints in the
more general theory with a gauge group $G\times U(1)$ and we
work out the Dirac bracket between the basic dynamical
variables. In contrast to the $U(1)$-case, the analysis can be
performed without breaking manifest covariance, by taking
advantage of the peculiar group structure.  Finally we discuss
the global charges arising when the spatial manifold has a
boundary, and show that they fulfill, in five dimensions, the
WZW$_4$ algebra found in Ref. \cite{Nair}. We summarize and
conclude in Sec. \ref{conclusions}.

\section{Local dynamics}
\setcounter{equation}{0}
\label{paper1}

\subsection{Conventions and definitions}

The Chern-Simons action in higher odd dimensions is a direct
generalization of the three dimensional case. Let us consider a
Lie algebra $G$ of dimension $N$. Let $\tilde F^a$ be the
curvature 2-form\footnote{We denote by $\tilde F^a=\frac{1}{2}
F^a_{\mu\nu}dx^\mu \mbox{\tiny $\wedge\,$} dx^\nu $ the {\em
spacetime} curvature 2-form (greek indices run over spacetime
while latin indices run over the spacelike hypersurfaces). The
symbol $F^a$ will denote the spacelike curvature 2-form, $F^a=
\frac{1}{2} F^a_{ij} dx^i \mbox{\tiny $\wedge\,$} dx^j$.}
$\tilde F^a = dA^a + \frac{1}{2} f^a_{\ bc} A^b \mbox{\tiny
$\wedge\,$} A^c$ associated to the gauge field 1-form $A^a$,
where $f^a_{\ bc}$ are the structure constants of the gauge
group, and let $g_{a_1\ldots a_{n+1}}$ be a rank $n+1$,
symmetric tensor invariant under the adjoint action of the gauge
group. The Chern-Simons Lagrangian in $D=2n+1$ dimensions ${\cal
L}^{2n+1}$ is defined through the formula
\begin{equation}
d{\cal L}^{2n+1}=g_{a_1\ldots a_{n+1}}
\tilde F^{a_1}\mbox{\tiny $\wedge\,$}\cdots\mbox{\tiny
$\wedge\,$}\tilde F^{a_{n+1}}. 
\label{l2n+1}
\end{equation}
The Chern-Simons action $I=\int_M {\cal L}_{CS}^{2n+1}$ is
invariant, up to a boundary term, under standard gauge
transformations
\begin{equation}
\delta_\epsilon A^a_\mu = -D_\mu \epsilon^a.
\label{gauge}
\end{equation}
It is also invariant under diffeomorphisms on the spacetime
manifold $M$, $\delta_\eta A^a_\mu = \pounds_\eta A^a_\mu$,
because ${\cal L}_{CS}^{2n+1}$ is a ($2n+1$)-form. The
diffeomorphisms on $M$ can be represented equivalently by
\begin{equation}
\delta_\eta A^a_\mu = -\eta^\nu F^a_{\mu \nu}.
\label{diff2}
\end{equation}
This transformation differs from the Lie derivative only by a
gauge transformation and it is often called improved
diffeomorphism \cite{impdiff}.

If the only symmetries of the Chern-Simons action are the
diffeomorphisms (\ref{diff2}) and the gauge transformations
(\ref{gauge}), then we shall say that there is no accidental
gauge symmetry.  How this translates into an algebraic condition
on the invariant tensor  $g_{a_1\ldots a_{n+1}}$ will be
described precisely in Sec.  \ref{Generic}. As we shall also
indicate, the absence of accidental gauge symmetries is
``generic''. Generic, however, does not mean universal and there
exist examples with further gauge symmetries. A typical one is
obtained by taking all the mixed components of $g_{a_1\ldots
a_{n+1}}$  equal to zero, so that the action is just the direct
sum of $N$ copies of the action for a single Abelian field.
This theory is then clearly invariant under diffeomorphisms
acting independently on each copy.  But there is no reason to
take vanishing mixed components for $g_{a_1\ldots a_{n+1}}$.  In
fact, for a non-Abelian theory, the form of the invariant tensor
is severely restricted. If the mixed components of $g_{a_1\ldots
a_{n+1}}$ differ from zero  (and cannot be brought to zero by a
change of basis), then the action is not invariant under
diffeomorphisms acting independently on each gauge field
component $A^a$, because the invariance of the cross terms
requires the diffeomorphism parameters for each copy to be
equal, thus gluing all of them together in a single symmetry.

The Chern-Simons equations of motion are easily found to be
\begin{equation}
g_{aa_1\ldots a_{n}} \tilde F^{a_1}\mbox{\tiny $\wedge\,$}\cdots
\mbox{\tiny $\wedge\,$} \tilde F^{a_{n}} = 0
\label{equmotion}
\end{equation}
and they reduce to $\tilde F^a = 0$ only in the
three-dimensional case (provided of course that $g_{ab}$ is
invertible).

\subsection{The Hamiltonian action}

In order to perform the Hamiltonian analysis, we assume that the
spacetime manifold $M$ has the topology $\Re \times \Sigma$,
where $\Sigma$ is a $2n$-dimensional manifold. In this section,
we will concentrate on the local properties of the theory, that
is, we will not analyse the special features that arise if
$\Sigma$ has a boundary or a non-trivial topology. The presence
of boundaries will be considered in Sec. \ref{WZW}. We decompose
the spacetime gauge field 1-form $A^a$ as $A^a_\mu
dx^\mu=A^a_0dt + A^a_i dx^i$ where the coordinate $t$ runs over
$\Re$ and the $x^i$ are coordinates on $\Sigma$.  Although there
is no spacetime metric  to give any meaning to expressions such
as timelike or spacelike, we will call time the coordinate $t$
and we will say that $\Sigma$ is a spacelike section as
shorthand expressions.

It is easy to see that the Chern-Simons action depends linearly
on the time derivative of $A^a_i$,
\begin{equation}
I = \int_\Re \int_\Sigma [l^i_a(A^b_j) \dot A^a_i - A^a_0 K_a ],
\label{I2}
\end{equation}
where $K_a$ is given by
\begin{equation}
K_a =- \frac{1}{2^n n}g_{a a_1 a_2 \ldots a_n} 
\epsilon^{i_1 \ldots i_{2n}} 
F^{a_1}_{i_1 i_2} \cdots F^{a_n}_{i_{2n-1} i_{2n}} .
\label{K/general}
\end{equation}

The explicit form of the function $l^i_a(A^b_j)$ appearing in
Eq. (\ref{I2}) is not needed here but only its ``exterior''
derivative in the space of spatial connections, which reads
\begin{eqnarray}
\Omega^{ij}_{ab} &\equiv&
\frac{\delta l^j_b}{\delta A^a_i}-\frac{\delta l^i_a}{\delta A^b_j}
\nonumber \\
  &=& -\frac{1}{2^{n-1}} \epsilon^{ij i_1 \ldots  i_{2n-2}} 
g_{ab a_1 \ldots a_{n-1}} F^{a_1}_{i_1 i_2} \cdots 
F^{a_{n-1}}_{i_{2n-3}i_{2n-2}}.
\label{Omega}
\end{eqnarray}

The equations of motion obtained by varying the action
(\ref{I2}) with respect to $A^a_i$ are given by
\begin{equation}
\Omega^{ij}_{ab}\dot A^b_j=\Omega^{ij}_{ab}D_jA^b_0 ,
\label{A-dot}
\end{equation}
while the variation of the action with respect to $A^a_0$ yields
the constraint
\begin{equation}
 K_a=0.
\label{K=0}
\end{equation}
Of course, Eqs. (\ref{A-dot}) and (\ref{K=0}) are completely
equivalent to Eq. (\ref{equmotion}). Despite  the fact that Eqs.
(\ref{A-dot}) are  first order, they are not Hamiltonian.  The
reason is that the matrix $\Omega^{ij}_{ab}$ is not invertible,
as we will see.   Indeed, on the surface defined by the
constraint (\ref{K=0}), the matrix $\Omega$ has, at least, $2n$
null eigenvectors.  The non-invertibility of $\Omega$ is  a
signal of a gauge symmetry. We shall see below that this
symmetry is nothing but the diffeomorphism invariance.

To proceed with the Hamiltonian formulation of the action we
shall use the Dirac method \cite{HT-book}.  Since the action
(\ref{I2}) is linear in the time derivatives of $A^a_i$, the
canonically conjugate momenta $p^i_a$ are subject to the $2n N$
primary constraints,
\begin{equation}
\phi^i_a = p^i_a - l^i_a \approx 0 \ .
\label{primary}
\end{equation}
These constraints transform in the coadjoint representation of
the Lie algebra because the inhomogeneous terms in the
transformation laws of $p_a^i$ and $l_a^i$ cancel out.

In principle one should also define a canonical momentum for
$A^a_0$.  This would generate another constraint, $p^0_a=0$,
whose consistency condition yields the constraint $K_a=0$. The
constraint $p^0_a=0$ is  first class and generates arbitrary
displacements of $A^a_0$. One can drop $p^a_0$ and keep $A^a_0$
as an arbitrary Lagrange multiplier for the constraint $K_a=0$.
 
It turns out to be more convenient to replace the constraints
$K_a$ by the equivalent set
\begin{equation}
G_a=-K_a+D_i\phi^i_a.
\label{G}
\end{equation}
This redefinition is permissible because the surface defined by
$K_a=0$, $\phi^a_i =0$ is equivalent to the surface defined by
$G_a=0$, $ \phi^a_i =0$.   The motivation to replace $K_a$ by
$G_a$ is that the new constraints $G_a$ generate the gauge
transformations (\ref{gauge}) and therefore are first class.
Indeed one can easily check that,
\begin{equation}
\delta A^a_i=\{A_i^a,\int_\Sigma\lambda^b G_b\}=-D_i\lambda^a.
\end{equation}

The Hamiltonian action  takes the form,
\begin{equation}
I= \int_\Re \int_\Sigma [ p^i_a \dot A^a_i - A^a_0 G_a - u^a_i
\phi^i_a ],
\end{equation}
where the Poisson brackets among the constraints are given by
\begin{eqnarray}
\{ \phi^i_a , \phi^j_b \} & =& \Omega^{ij}_{ab},\label{phi-phi} \\
\{\phi^i_a, G_b\} &=&f^c_{\ ab} \phi^i_c ,\label{phi-G}\\
\{ G_a,G_b\} &=& f^c_{\ ab} G_c. \label{G-G}
\end{eqnarray}
Here,  $f^c_{\ ab}$ are the structure constants of the Lie
algebra under consideration.

It follows from the constraint algebra that there are no further
constraints. The consistency condition
\begin{equation}
\dot G_a=0
\label{G-dot}
\end{equation} 
is automatically fulfilled because $G_a$ is first class while
the other consistency equation
\begin{equation}
\dot\phi^i_a=\Omega^{ij}_{ab}u_j^b=0
\label{phi-dot}
\end{equation}
will just restrict some of the Lagrange multipliers $u^b_j$.

\subsection{First class constraints}

Equations (\ref{phi-G}) and (\ref{G-G}) reflect that the
constraints $G_a$  are the generators of the gauge
transformations and that the constraints $\phi_a^i$ transform in
the coadjoint representation.  This means, in particular, that
the $G_a$'s are first class, as mentioned above.

The nature of the constraints $\phi_a^i$ is determined by the
eigenvalues of the matrix $\Omega^{ij}_{ab}$. It turns out that
the matrix $\Omega^{ij}_{ab}$ is not invertible on the
constraint surface and, therefore, not all the $\phi$'s are
second class.  Indeed, using some simple combinatorial
identities, one can prove that $K_a$ and $\Omega^{ij}_{ab}$
satisfy the identity
\begin{equation}
\Omega^{ij}_{ab} F^b_{kj} = \delta^i_k K_a.
\label{null-diff}
\end{equation}
This equation shows that, on the constraint surface $K_a=0$, the
matrix $\Omega^{ij}_{ab}$ has, at least, $2n$ null eigenvectors
$(v_k)^b_j=F^b_{kj},\ (k=1\ldots 2n)$.  The existence of these
$2n$ null eigenvectors of $\Omega$ tells us that among the
$\phi$'s, there are $2n$ first class constraints. These
constraints are given by
\begin{equation}
H_i \equiv F^a_{ij} \phi^j_a,
\label{Hi}
\end{equation}
and they generate the spatial diffeomorphisms (\ref{diff2}),
namely, $\delta A^a_i=\{A_i^a,\int_\Sigma\eta^j H_j\}= -\eta^j
F_{ij}$.  Thus, they satisfy the spatial diffeomorphism algebra,
up to gauge transformations. The presence of these constraints
is not surprising because the Chern-Simons action is invariant
under diffeomorphisms for any choice of the invariant tensor
$g_{a_1\ldots a_{n+1}}$.  What is perhaps more surprising in
view of what occurs for ordinary Chern-Simons theory in three
dimensions, is that the constraints (\ref{Hi}) are generically
independent from the constraints  $G_a= 0$ generating local
internal gauge transformations ($F^a_{ij} \not= 0$, see below).

One could also expect the presence of another first class
constraint, namely, the generator of timelike diffeomorphisms.
However, this symmetry is not independent from the other ones
and hence its generator is a combination of the first class
constraints $G_a$ and $H_i$.  This can be viewed as follows. The
action of a timelike diffeomorphism parameterized by
$\xi^\mu=(\xi^0,0)$ acting on $A^a_i$ is [see Eq.
(\ref{diff2})],
\begin{equation}
\delta_\xi A^a_i=-\xi^0F_{i0}^a.
\label{tdiff}
\end{equation}
Now, the equations of motion (\ref{A-dot}) are
$\Omega^{ij}_{ab}F^b_{0j}=0$. Let us assume (this assumption
will be clarified below) that the only null eigenvectors of
$\Omega$ are those given above, then there must exist some
$\zeta^k$ such that $F^b_{j0}=\zeta^k F^b_{jk}$.  Inserting this
result in Eq. (\ref{tdiff}), we obtain
\begin{equation}
\delta_\xi A^a_i= -\xi^0\zeta^kF_{ik}^a,
\end{equation}
which is a spatial diffeomorphism with parameter $\xi^0\zeta^k$.
Hence, timelike diffeomorphisms are equivalent to spacelike
diffeomorphisms on-shell. For that reason, there is no
constraint associated to normal deformations of the surface.  If
the theory has `accidental symmetries', i.e. the matrix $\Omega$
has more zero eigenvalues, the above analysis shows that the
timelike diffeomorphisms can still be written in terms of
spatial diffeomorphisms plus the extra `accidental' symmetries.

\subsection{Generic theories}
\label{Generic}

We now examine whether the first class constraints $G_a$ and
$H_i$ are independent and constitute a complete set. At this
point, we must distinguish between the $N=1$ and $N>1$ cases.
Indeed, it turns out that for the Lie algebra $u(1)$, the theory
cannot be obtained as a limiting case from the $N>1$ theory.

Consider first the general case with $N>1$. The eigenvalues of
$\Omega$, which determine the nature of the constraints
$\phi^i_a$, depend on the properties of the invariant tensor
$g_{a_1\ldots a_{n+1}}$ and, for  a definite choice of
$g_{a_1\ldots a_{n+1}}$,  they  also depend on the phase space
location of the system since the constraint surface of the
Chern-Simons theory is stratified into phase space regions where
the matrix $\Omega_{ab}^{ij}$ has different ranks.  For example,
$F^a=0$ is always a solution of the equations of motion and, for
that solution, $\Omega$ is identically zero.  There exist,
however, other solutions of the equations of motion for which
$\Omega\neq 0$. The rank of $\Omega$ classifies the phase space
into regions with different number of local  degrees of freedom.
The key ingredient controlling the maximum possible rank of
$\Omega$ is the algebraic structure of the invariant tensor
$g_{a_1\ldots a_{n+1}}$.

We will say that an invariant tensor $g_{a_1\ldots a_{n+1}}$ is {\em
generic} if and only if it satisfies the following condition:
There exist solutions $F^a_{ij}$ of the constraints $K_a=0$ such
that
\begin{enumerate}
\item[ $(i)$]  the matrix $F^b_{kj}$ (with $b,j$ as row index and
$k$ as column index) has maximum rank $2n$, so that the only
solution of $\xi^kF^b_{kj}=0$ is $\xi^k=0$ and therefore the
$2n$ null eigenvectors $(v_k)^b_j=F^b_{kj},\ (k=1\ldots 2n)$ are
linearly independent;

\item[$(ii)$] the $(2nN)\times (2nN)$ matrix $\Omega_{ab}^{ij}$
has the maximum rank compatible with $(i)$, namely $2nN-2n$; in
other words, it has no other null eigenvectors besides
$(v_k)^b_j=F^b_{kj},\ (k=1\ldots 2n)$.
\end{enumerate}
We will also say that  the solutions $F^a_{ij}$ of the
constraints $K_a=0$ such that $(i)$ and $(ii)$ hold are {\em
generic}.  The reason for this name comes from the following
observation. For a given generic tensor $g_{a_1\ldots  a_{n+1}}$, a
solution fulfilling both conditions $(i)$ and $(ii)$ will still
fulfill them upon small perturbations, since maximum rank
conditions correspond to inequalities and therefore, they define
open regions.  Conversely, a solution not fulfilling conditions
$(i)$ or $(ii)$, i.e. located on the surface where lower ranks
are achieved (defined by equations expressing that some non
trivial determinants vanish), will fail to remain on that
surface upon generic perturbations consistent with the
constraints.  Thus, non generic solutions belong to subsets of
the constraint surface of smaller dimension.

The physical meaning of the above algebraic conditions is
straightforward.  They simply express that the gauge
transformations (\ref{gauge}) and the spatial diffeomorphisms
(\ref{diff2}) are independent  and that the $H_i$ are the only
first class constraints among the $\phi^j_a$'s.

As we stressed above, the $N=1$ theory cannot be obtained as a
limiting case from the $N>1$ theory. The definition of what is
meant by ``generic" in the $N=1$ case must therefore be amended
as follows.  For $N=1$, the invariant tensor has a single
component $g_{1 \ldots 1}$, which we assume, of course, to be
different from zero.  The constraint  is then
\begin{equation}
\epsilon^{i_1\ldots i_{2n}} F_{i_1 i_2} \cdots F_{i_{2n-1} i_{2n}}=0 ,
\label{K/N=1}
\end{equation}
which implies that the matrix $F_{ij}$ cannot be invertible. The
solutions of Eq. (\ref{K/N=1}) are the set of matrices $F_{ij}$
with zero determinant.  Thus, the equation $\xi^i F_{ij}=0$ does
not imply $\xi^i=0$ and the constraint (\ref{K/N=1}) prevents 
us from finding solutions of the $N=1$ theory satisfying the
generic condition of the $N>1$ case. This means, in particular,
that the $2n$ spatial diffeomorphisms are not all independent.
Solutions of Eq. (\ref{K/N=1}) such that $F_{ij}$ has the
maximum rank compatible with the constraint (\ref{K/N=1}) will
be called  {\em generic}.  This rank is clearly $2n-2$.  The
complete Hamiltonian analysis for the $N=1$ theory is performed
in Sec.
\ref{N=1}.

\subsection{Degrees of freedom count  $(N>1)$}

When the $generic$ condition is satisfied, the count of local
degrees of freedom goes as follows.  We have, $2\times 2nN$
canonical variables ($A^a_i, p^i_a$), $N$ first class
constraints $G_a$ associated with the gauge invariance, $2n$
first class constraints $H_i$ associated with the spatial
diffeomorphism invariance, and $2nN-2n$ second class constraints
(the remaining $\phi^i_a$).  Hence, we have
\begin{eqnarray}
{\cal N}&=& \frac{1}{2}[4nN - 2(N + 2n) - (2nN - 2n)]
\nonumber\\ & = & nN-n-N
\label{DF}
\end{eqnarray}
local degrees of freedom (for $n>1$, $ N>1$).   It should be
stressed here that this formula gives the number of local
degrees of freedom associated to the open region of phase space
defined by generic solutions.

This formula does not apply to $N=1$ because the spatial
diffeomorphisms are not independent in that case.  One finds
instead that there are no local degrees of freedom (see Sec.
\ref{N=1}). For a similar reason, this formula does not apply to
$n=1$ ($D=3$) where  diffeomorphism invariance is completely
contained within the ordinary Yang-Mills gauge invariance.

\subsection{The generic condition in the Lagrangian 
equations of motion}

It is instructive to study the implications of the generic
condition in the context of the Lagrangian equations of motion.
This provides also an equivalent method of counting the number
of local degrees of freedom.  The Lagrangian equations of motion
written  in a covariant way are given in Eq. (\ref{equmotion}).
Upon a $(2n+1)$-decomposition, these equations acquire the
expressions given  in Eqs.  (\ref{A-dot}) and (\ref{K=0}).

If the generic condition is fulfilled, then Eq. (\ref{A-dot})
can be rewritten in the useful and simple form,
\begin{equation}
\dot A^a_i = D_i A^a_0 + N^k F^a_{ki},
\label{A-dot2}
\end{equation}
where the $N^k$ are $2n$ arbitrary functions of spacetime. This
form of the equations of motion clearly shows that the time
evolution is generated by a gauge transformation (with parameter
$A^a_0$) plus a diffeomorphism (with parameter $N^l$).  [Eq.
(\ref{A-dot2}) follows directly from Eq. (\ref{A-dot}), and the
fact that, in the generic case, $\Omega$ has only $2n$ null
eigenvectors given in Eq. (\ref{null-diff}).] Therefore, the
Lagrangian  equations of motion can be replaced by the
constraint (\ref{K=0}) plus Eq. (\ref{A-dot2}).  Due to the
simplicity of Eq. (\ref{A-dot2}) we can study its space of
solutions, modulo gauge transformations.

Equation (\ref{A-dot2}) is invariant under standard gauge
transformations,
\begin{equation}
\delta_\lambda A^a_i = -D_i \lambda^a, 
\hspace{7mm}
\delta_\lambda A^a_0 = -\dot \lambda^a - [\lambda,A_0]^a, 
\hspace{7mm}
\delta_\lambda N^k=0,
\label{delta-A0}
\end{equation}
where $[\cdot,\cdot]$ is the commutator in the Lie algebra.  Eq.
(\ref{A-dot2}) is also invariant under spatial diffeomorphisms
\begin{equation}
\delta_\xi A^a_i   = -\xi^j F^a_{ij},  
\hspace{7mm}
\delta_\xi A^a_0=-\xi^j F^a_{0j}, 
\hspace{7mm}
\delta_\xi N^i = \dot \xi^i + [\xi,N]^i,
\label{delta-N}
\end{equation}
where now the symbol $[\cdot,\cdot]$ denotes the Lie bracket of
two spatial vectors. Of course, the constraint equation
(\ref{K=0}) is also invariant under both  gauge transformations
and diffeomorphisms.

We shall study the solutions of the equations of motion in the
gauge
\begin{equation}
A^a_0 =0, \hspace{7mm} N^k=0, \hspace{7mm} (\mbox{time gauge}).
\label{time-gauge}
\end{equation}
In this gauge, Eq. (\ref{A-dot2}) simply says that $\dot
A^a_i=0$, hence, the configurations are time independent. It is
important to note that the above gauge choice does not exhaust
all the gauge symmetry.  The conditions (\ref{time-gauge}) are
preserved by gauge transformations and diffeomorphism that do
not depend on time, i.e., by transformations whose parameters
satisfy $\dot \lambda^a=0$ and $\dot \xi^i=0$.

Thus, in the time gauge, we are left only with the constraint
equation (\ref{K=0}) with the extra condition that the fields
are time independent.  Equation (\ref{K=0}) is invariant under
the residual gauge group consisting in time-independent gauge
transformations and diffeomorphisms. In summary, we have $2nN$
arbitrary functions of the spatial coordinates, $A^a_i(x^i)$.
These functions are restricted by $N$ equations, the constraints
(\ref{K=0}). Also, there is a $(N+2n)$-dimensional  residual
gauge group which can be used to set $N+2n$ functions equal to
zero.

Therefore, the number of arbitrary functions in the solutions of
the equations of motion is
\begin{equation}
2nN - N - (N+2n) = 2(nN-N-n).
\label{LDF}
\end{equation}
These are the Lagrangian ``integration functions" for the
equations of motion,  which are twice the number of local
degrees of freedom, in agreement with Eq. (\ref{DF}).

\section{Lovelock-Chern-Simons gravity} 
\setcounter{equation}{0}
\label{Gravity}

The goal of this section is twofold. On the one hand, we exhibit
the Lovelock-Chern-Simons theory as a concrete example with a
non abelian internal group for which the generic condition
defined in Sec. \ref{Generic} is fulfilled. On the other hand,
the analysis of the dynamics of the Lovelock-Chern-Simons action
reveals a rather unexpected result.  It turns out that  the
torsion tensor is dynamical in this theory, hence the Palatini
and second order formalisms are not equivalent in higher
dimensional Chern-Simons gravity, contrary to what happens in
three dimensions.

The Lovelock-Chern-Simons theory (in higher dimensions) is not
defined by the Hilbert-Einstein action but, rather, it contains
higher powers of the curvature tensor. However, the equations of
motion are  first order in the tetrad and spin connection, and,
if the torsion is set equal to zero, they are  second order in
the metric.  The gravitational Chern-Simons action is a
particular case of the so-called Lovelock action
\cite{Lovelock}.  For this reason, we  call this theory the
Lovelock-Chern-Simons (LCS) theory. It is a natural extension to
higher dimensions of the formulation of 2+1 gravity given by
Ach\'ucarro and Townsend \cite{Achucarro-Townsend} and Witten
\cite{Witten88}.  As we shall see, however, the dynamical
content is quite different.  The construction of the
Lovelock-Chern-Simons theory has been carried out in Ref.
\cite{Chamseddine}. Here we briefly review its main features.

Let $A^{AB}$ be a connection for the group $SO(2n,2)$ [we recall
that $D=2n+1$] and $F^{AB}$ its curvature 2-form (here, the
capital indices $A,B\ldots $ run over $1,2\ldots 2n+2$).  The
connection $A^{AB}$ can be split in the form,
\begin{equation}
A^{AB} =  \left(
\begin{array}{cc} w^{ab} & e^a/l \\ 
               - e^b/l   &  0    
\end{array}  \right),
\label{W-ads} 
\end{equation}
where $l$ parameterizes the cosmological constant and $w^{ab}$
and $e^a$ transform, respectively, as a connection and as a
vector under the action of the Lorentz subgroup $SO(2n,1)$.
Hence, $w^{ab}$ will be called the spin connection and $e^a$ the
vielbein.   Similarly, the curvature has the form
\begin{equation}
F^{AB} =  \left(
\begin{array}{cc} F^{ab}   &  T^a/l \\ 
                 - T^b/l   &  0    
\end{array}  \right),
\label{R-ads} 
\end{equation}
where $F^{ab}=R^{ab}+\frac{1}{l^2} e^a \mbox{\tiny $\wedge\,$}
e^b$ and $T^a=De^a$ is the torsion tensor.

The $SO(2n,2)$ Chern-Simons Lagrangian is defined by making use
of the Levi-Civita invariant tensor,
\begin{equation}
d{\cal L}_{LCS} = \epsilon_{A_1\ldots A_{2n+2}} 
F^{A_1 A_2} \mbox{\tiny $\wedge\,$}
\cdots \mbox{\tiny $\wedge\,$} F^{A_{2n+1} A_{2n+2}}.
\label{LCS}
\end{equation} 
Since $\epsilon_{A_1\ldots A_{2n+2}}$ is an invariant tensor of
$SO(2n,2)$, we say that ${\cal L}_{LCS}$ is a Chern-Simons
Lagrangian for the (adS) group $SO(2n,2)$.  When written in
terms of the vielbein and spin connection, the Lagrangian
defined in Eq. (\ref{LCS}) is a particular case of the Lovelock
Lagrangian considered in Ref. \cite{Lovelock}. Black hole
solutions for this action have been found in Ref. \cite{BTZ2}.

The equations of motion for this theory are
\begin{equation}
\epsilon_{ABA_1\ldots A_{2n}} 
F^{A_1 A_2} \mbox{\tiny $\wedge\,$} \cdots 
\mbox{\tiny $\wedge\,$} F^{A_{2n-1} A_{2n}} =0,
\label{W}
\end{equation}
which are explicitly invariant under $SO(2n,2)$.  Splitting the
curvature $F^{AB}$ as in Eq. (\ref{R-ads}), these equations
separate into the two sets of equations,
\begin{eqnarray}
\epsilon_{aa_1\ldots a_{2n}} F^{a_1 a_2} 
\mbox{\tiny $\wedge\,$} \cdots \mbox{\tiny $\wedge\,$} 
F^{a_{2n-1} a_{2n}} =0,  \label{e}   \\
\epsilon_{aba_1\ldots a_{2n-1}} 
F^{a_1 a_2} \mbox{\tiny $\wedge\,$} \cdots 
\mbox{\tiny $\wedge\,$} F^{a_{2n-3} a_{2n-2}} 
\mbox{\tiny $\wedge\,$} T^{a_{2n-1}}  =0, \label{w}
\end{eqnarray}
which are the equations of motion following from varying the
action with respect to the vielbein and spin connection,
respectively.  After the split has been made, the equations  are
explicitly invariant only under $SO(2n,1)$ but, in view of Eq.
(\ref{W}), they are in fact invariant under the larger group
$SO(2n,2)$. Note that, as mentioned above, $T^a=0$ is always a
solution of Eq. (\ref{w}) and, in $2+1$ dimensions, it is the
only solution. We now prove that, for $n>1$, this is {\em not}
the most general solution and {\em dynamical} $T^a\neq 0$ modes
exist.

For simplicity, we consider the five dimensional case. As it has
been shown in Ref. \cite{TZ}, the Lovelock Lagrangian written in
the second order formalism ($T^a=0$) has the same number of
degrees of freedom as the Hilbert Lagrangian.  Thus, in five
dimensions, the theory with zero torsion carries $D(D-3)/2=5$
local (physical) degrees of freedom.  On the other hand, since
the Lagrangian ${\cal L}_{LCS}$ defines a Chern-Simons theory we
can count the number of degrees of freedom by using the formula
(\ref{DF}). However, before we can apply that formula we need to
prove that the Lovelock-Chern-Simons theory is generic in the
sense defined in Sec. \ref{Generic}.

The constraint for this theory is
\begin{equation}
\epsilon_{ABCDEF} F^{CD} \mbox{\tiny $\wedge\,$} F^{EF}=0,
\label{K/gravity}
\end{equation}
where $F^{CD} = \frac{1}{2} F^{CD}_{ij} dx^i \mbox{\tiny
$\wedge\,$} dx^j $ are the spatial projections of the 2-form
curvature. To prove that this theory is generic, it is enough to
find one solution for which the matrix $\Omega$ has maximum
rank. The 2-form curvature given by
\begin{eqnarray}
F^{12} &=& dx^1 \mbox{\tiny $\wedge\,$} dx^2 + 
dx^3 \mbox{\tiny $\wedge\,$} dx^4, \nonumber\\ 
F^{34} &=& dx^1 \mbox{\tiny $\wedge\,$} dx^2 - 
dx^3 \mbox{\tiny $\wedge\,$} dx^4, \label{grav/F} \\ 
F^{56} &=& dx^1 \mbox{\tiny $\wedge\,$} dx^3 + 
dx^2 \mbox{\tiny $\wedge\,$} dx^4 . \nonumber
\end{eqnarray}
with all other components equal to zero satisfies the constraint
(\ref{K/gravity}) and has maximum rank. The proof of this
statement is straightforward. One looks at the equation
\begin{equation}
\epsilon_{ABCDEF} F^{CD} \mbox{\tiny $\wedge\,$} V^{EF}=0,
\end{equation}
 where $V^{EF}$ is a one form. Due to the fact that
$F^{12},F^{34}$ and $F^{56}$ are non-degenerate 2-forms, one
easily obtains that this equation possesses only four
independent solutions, that is, $\Omega$ has the maximum rank.

Note that $F^{AB}$ given in Eq. (\ref{grav/F}) does not have any
zero column in the indices $(A,B)$ therefore this solution
clearly has a non-zero torsion [see Eq. (\ref{R-ads})]. Note
also that the above curvature can be derived from the connection
$W^{AB}$
\begin{eqnarray}
W^{12} &=& x^1 dx^2 + x^3 dx^4, \nonumber\\ 
W^{34} &=& x^1 dx^2 - x^3 dx^4,\\ 
W^{56} &=& x^1 dx^3 + x^2 dx^4 . \nonumber
\end{eqnarray}
with all other components equal to zero. Thus, $F^{AB}$ given in
Eq. (\ref{grav/F}) represents an allowed physical configuration.

The existence of the above solution ensures that this theory is
generic and therefore we can apply the formula (\ref{DF}) to
count the number of degrees of freedom.  In five dimensions, the
Lovelock-Chern-Simons theory is a Chern-Simons theory for the
Lie algebra $SO(4,2)$, of dimension 15. Hence, formula
(\ref{DF}) gives $2\times 15-15-2=13$ local degrees of freedom.
Thus, this theory indeed has more degrees of freedom than the
metric theory. This reflects the fact that setting the torsion
equal to zero eliminates degrees of freedom.  In other words,
Lovelock theory, at least in the case considered here, has a
dynamical torsion.

\section{The seven dimensional case}
\setcounter{equation}{0}
\label{D=7}

In Ref. \cite{BGH}, examples fulfilling the generic condition
were explicitly given only in five dimensions.  We exhibit in
this section seven dimensional examples for which the generic
condition is satisfied.  In this case, there exists a `simple'
choice for the invariant tensor: We can take the rank-four
invariant symmetric tensor $g_{abcd}$ given by
\begin{equation}
g_{abcd} = g_{ab}g_{cd} + g_{ac}g_{bd} + g_{ad}g_{bc},
\label{g/7}
\end{equation}
where $g_{ab}$ is an invariant metric on the Lie algebra. We
prove in this section that if $g_{ab}$ is invertible, then the
associated Chern-Simons theory is generic.  This means, in
particular, that the seven dimensional theory with the choice
(\ref{g/7})  --- or with any other choice of invariant tensor
sufficiently close to it --- is generic for any simple Lie
algebra.
 
The constraint in this case reduces to the simpler form
\begin{equation}
K_a=-F_a \mbox{\tiny $\wedge\,$} 
F^b \mbox{\tiny $\wedge\,$} F_b = 0,
\label{K/7}
\end{equation}
where the internal indices are raised and lowered with $g_{ab}$.
Here $F^a=\frac{1}{2}F^a_{ij}dx^i\mbox{\tiny $\wedge\,$} dx^j$,
and we regard the constraint as a 6-form.  Similarly, 
$\Omega$ is a 4-form given by,
\begin{equation}
\Omega_{ab} = - g_{ab} F^c \mbox{\tiny $\wedge\,$} 
F_c -2  F_a \mbox{\tiny $\wedge\,$} F_b. 
\label{O}
\end{equation}
We want to find solutions to Eq. (\ref{K/7}) so that $\Omega$
has maximum rank.  Thus, we are interested in the number of
solutions of the null eigenvalue problem
\begin{equation}
\Omega_{ab} \mbox{\tiny $\wedge\,$} V^b = 0 ,
\label{V/7}
\end{equation}
where $V^b$ is a 1-form vector, and $F^a$ satisfies Eq.
(\ref{K/7}).  We already know that this equation has six
independent solutions which are of the form
$V^a_i=F^a_{ij}\xi^j$. We will now show that there exists a
solution to the constraint (\ref{K/7}) for which $\Omega$ does
not have any other null eigenvectors.

The constraint (\ref{K/7}) is solved by the following expression
for $F^a$:
\begin{equation}
F^a= f^a dx^{1}\mbox{\tiny $\wedge\,$} dx^{2} + 
g^a dx^{3}\mbox{\tiny $\wedge\,$} dx^{4} + 
h^a dx^{5}\mbox{\tiny $\wedge\,$} dx^{6},
\label{F-sol}
\end{equation}
where
\begin{equation}
f^a = x^a + \sqrt{3} y^a, 
\hspace{7mm} 
g^a = x^a - \sqrt{3} y^a, 
\hspace{7mm}
h^a = x^a 
\end{equation}
and $x^a$, $y^a$ are two vectors satisfying $x^a x_a=1=y^ay_a$,
$x^ay_a=0$. Here we have assumed that $g_{ab}=\delta_{ab}$ only
for simplicity. The analysis can be carried out for any
invertible $g_{ab}$.

The matrix $\Omega$ evaluated for this solution is equal to
\begin{eqnarray}
\Omega_{ab} = 2\hspace{-5mm} &&\left( A_{ab}\ dx^{1}
\mbox{\tiny $\wedge\,$} dx^{2}
\mbox{\tiny $\wedge\,$} dx^{3}\mbox{\tiny $\wedge\,$} dx^{4}
\right.
\nonumber\\
&&+ \left.  B_{ab}\ dx^{1}\mbox{\tiny $\wedge\,$} dx^{2}
\mbox{\tiny $\wedge\,$} dx^{5}\mbox{\tiny $\wedge\,$} dx^{6} 
\right.
\\
&&+  \left. C_{ab}\ dx^{3}\mbox{\tiny $\wedge\,$} dx^{4}
\mbox{\tiny $\wedge\,$} dx^{5}\mbox{\tiny $\wedge\,$} dx^{6}\right),
\nonumber
\end{eqnarray}
where 
\begin{eqnarray}
A_{ab} &=& (f^c g_c)  g_{ab} + f_a g_b + f_b g_a,  \nonumber\\
B_{ab} &=& (f^c h_c)  g_{ab} + f_a h_b + f_b h_a, \label{abc} \\
C_{ab} &=& (h^c g_c)  g_{ab} + h_a g_b + h_b g_a.\nonumber
\end{eqnarray}

An immediate set of null eigenvectors of $\Omega$ comes from the
observation that $h^a,g^a$ and $f^a$ are, respectively, null
eigenvectors of $A,B$ and $C$. These eigenvectors are easily
seen to correspond to the diffeomorphisms eigenvectors $\xi^i
F^a_{ij}$.

To prove that this theory has maximum rank, it  is now enough to
prove that the matrices $A,B$ and $C$ do not have any further
null eigenvectors.  This is most easily shown by going to the
particular basis in which,
\begin{equation}
x^a = (1,0,0,\ldots ,0), \hspace{7mm} y^a = (0,1,0,\ldots ,0).
\end{equation}
Then, the vectors $f,g$ and $h$ have the form
\begin{equation}
f^a = (1, \sqrt{3}, 0 , \ldots  ,0), \hspace{7mm}
g^a = (1, -\sqrt{3}, 0 , \ldots  ,0), \hspace{7mm}
h^a = (1,0 , 0 , \ldots  , 0)
\end{equation}
and $A$, $B$ and $C$ have the block form
\begin{eqnarray}
A &=& \left( \begin{array}{cc|c}   
0 &  0  &  0  \\ 
0  &  -8  &0  \\ \hline
0  &  0   &  -2 I  
 \end{array}  \right), \nonumber\\
B &=&  \left( \begin{array}{cc|c}  
 3  &  \sqrt{3} & 0 \\
  \sqrt{3}   &  1 &0  \\ \hline 
0&  0   &  I    
\end{array}  \right), \\
C &=&  \left(   \begin{array}{cc|c} 
 3  &  -\sqrt{3} &0  \\ 
  -\sqrt{3}   &  1 & 0  \\ \hline   
0 & 0   &  I    
\end{array}  \right),\nonumber
\end{eqnarray}
where $I$ is the identity in the $(N-2)\times (N-2)$ subspace
orthogonal to $x^a$ and $y^a$. Accordingly, $A,B$ and $C$ have
rank $N-1$ showing that each of them has only one null
eigenvector.  These eigenvectors are, respectively, $h^a$, $g^a$
and $f^a$.

Thus, we have proved that the seven dimensional Chern-Simons
theory defined by the invariant tensor given in Eq. (\ref{g/7})
provides another example of generic theories.

\section{The $(N=1)$ Abelian theory}
\setcounter{equation}{0}
\label{N=1}

The $N=1$ theory was first studied in Ref. \cite{Floreanini}
were the absence of degrees of freedom for this theory was
pointed out.  Here we shall analyse this theory along the lines
introduced in Sec. \ref{paper1}. Our main goal is to display the
differences between the $N=1$ and $N>1$ theories.

As it was pointed out at the end of Sec. \ref{Generic}, the
$N=1$ theory needs a special treatment. We defined in that
section the generic solutions of the constraint as those for
which the matrix $F_{ij}$ has the maximum possible rank $2n-2$.

Let us split the $2n$ spatial coordinates $x^i$ into
$(x^{\alpha}, x^p)$ where $x^\alpha=(x^1,x^2)$ and
$x^p=(x^3,x^4,\ldots,x^{2n})$, so that the gauge field takes the
form
\begin{equation}
A_i dx^i = A_\alpha dx^\alpha + A_p dx^p
\end{equation}
and the curvature can be written as
\begin{equation}
F_{ij} dx^i\mbox{\tiny $\wedge\,$} dx^j =
F_{\alpha\beta}dx^\alpha \mbox{\tiny $\wedge\,$}
dx^\beta + 2 F_{\alpha p}dx^\alpha \mbox{\tiny $\wedge\,$} dx^p 
+ F_{pq} dx^p \mbox{\tiny $\wedge\,$} dx^q.
\label{F/1}
\end{equation}
The generic condition for the $N=1$ theory can be implemented by
requiring that the $(2n-2)\times (2n-2)$ matrix $F_{pq}$
appearing in Eq. (\ref{F/1}) is invertible, i.e.
\begin{equation}
\det{F_{pq}} \neq 0,
\label{det F}
\end{equation}
so that $F_{ij}$ has the maximum rank $2n-2$.  By a change of
coordinates, one can always make a generic $F_{ij}$ to fulfill
the condition (\ref{det F}).

\subsection{Dirac brackets and first class algebra}

Once the maximum rank condition over $F_{pq}$ is imposed, the
constraints $\phi^i=(\phi^\alpha,\phi^p)$ split naturally into
first and second class.  To see this, we first note that
$\Omega^{\alpha \beta}$ (the projection of $\Omega^{ij}$ along
the coordinates $x^\alpha$) can be written as
\begin{equation}
\Omega^{\alpha\beta} = -\frac{1}{2^{n-1}}\epsilon^{\alpha\beta} f
\end{equation}
where $f$ is the Pfaffian of $F_{pq}$
\begin{equation}
f \equiv \epsilon^{p_1\cdots p_{2n-2}} F_{p_1p_2}\cdots
F_{p_{2n-3}p_{2n-2}}=\sqrt{\det{F_{pq}}} 
\label{pf}
\end{equation}
and $\epsilon^{\alpha \beta}$ is the Levi-Civita tensor in the
2-dimensional manifold labeled by the coordinates $x^\alpha$.
Since the determinant of $F_{pq}$ is different from zero, the
$2\times 2$ matrix $\Omega^{\alpha\beta}$ is invertible.  Let
$J$  be its inverse:
\begin{equation}
J_{\alpha \beta} = \frac{2^{n-1}}{f}\epsilon_{\alpha \beta},
\label{J/1}
\end{equation}
which satisfies $J_{\alpha \beta} \Omega^{\beta \gamma} =
\delta^\gamma_\alpha$.

The invertibility of $\Omega^{\alpha\beta}$ implies, from Eq.
(\ref{phi-phi}), that the two constraints $\phi^{\alpha}$ are
second class.  Consequently, their associated Lagrange
multipliers $u_\alpha$ can be solved from Eq. (\ref{phi-dot}),
\begin{equation}
u_\alpha =  J_{\alpha \beta} \Omega^{\beta p} u_p \ .
\label{ualpha}
\end{equation}

As we have seen, one may take as first class constraints the
$2n$ combinations  $H_i=F_{ij} \phi^j$.  In the $N=1$ case,
there are only $2n-2$ independent constraints among the $H_i$'s,
since the matrix $F_{ij}$ is of rank $2n-2$. If we recall that
the matrix $F_{pq}$ is invertible, we can take the independent
first class constraints to be
$H_p=F_{pq}\phi^q+F_{p\alpha}\phi^\alpha$.  The system of
constraints ($\phi^\alpha , H_p$) provides a system equivalent
to the system ($\phi^j$), in which the  constraints are
manifestly split into second and first class. Upon elimination
of the second class constraints, we obtain the corresponding
Dirac bracket
\begin{equation}
\{ A, B \}^* = \{A, B \} -  \int_\Sigma dz \{A, \phi^\alpha(z)\}
J_{\alpha \beta}(z) \{\phi^\beta(z), B \} .
\label{Dirac-bracket/1}
\end{equation}

The smeared generators $G(\lambda) = \int_\Sigma \lambda  G$ and
$\bar H(\xi) = \int_\Sigma \xi^p \bar H_p$, where $\bar
H_p=F_{pq}\phi^q+A_p G$ satisfy the Dirac bracket algebra,
\begin{eqnarray}
\{ G(\lambda), G(\eta) \}^* &=& 0,  
\label{G-G/1} \\
\{ \bar H(\xi) , G(\lambda) \}^* &=& G(\xi^p \partial _p \lambda),
\label{H-G} \\
\{ \bar H(\xi) , \bar H(\zeta) \}^* &=& \bar H([\xi,\zeta]), 
\end{eqnarray}
where $[\xi,\zeta]^p = \xi^q \zeta^p,_q - \zeta^q \xi^p,_q$ is
the Lie bracket of the two vectors $\xi^p$ and $\zeta^q$. The
above algebra is self explanatory; $G$ generates gauge
transformations and $\bar H_p$ generates diffeomorphisms in the
$x^p$ directions.  Equation (\ref{H-G}), on the other hand,
tells us that $G$ transforms as a scalar under diffeomorphisms.
 
This completes the problem of relating the first class algebra
with the symmetries of the Chern-Simons  action.  A different
question is whether the above constraints encode all the
symmetries of the action.  Two evident missing pieces are
diffeomorphisms along the $x^\alpha$ directions and timelike
diffeomorphisms.  It  turns out that these symmetries are
generated by the above constraints.  The proof of this statement
is straightforward.  The key point is the fact that the
equations of motion trivialize the timelike (as in the case
$N>1$) and $x^\alpha$-diffeomorphisms.  That is, when acting on
the space of solutions of the equations of motion, these
symmetries reduce to the identity. For this reason, there are no
independent first class constraints associated to them.  In a
canonical language, the generators of timelike and
$x^\alpha$-diffeomorphisms are linear combinations of the
generators of gauge transformations and $x^p$-diffeomorphisms.
This is reminiscent of the 2+1 theory where the whole
diffeomorphism invariance can be expressed in terms of the local
gauge transformations.

There is thus a gradation that can be summarized as follows: in
three dimensions, the diffeomorphisms can be expressed in terms
of the internal gauge transformations for any choice of the
gauge group.  In higher dimensions, some of the diffeomorphisms
become independent gauge symmetries (in the generic case). These
are $2n-2$ of the spatial diffeomorphisms if the gauge group is
one-dimensional, and all the spatial diffeomorphisms otherwise.
The timelike diffeomorphisms are not independent gauge
symmetries in any (generic) case; they can always be expressed
in terms of the other symmetries.

\subsection{Absence of local degrees of freedom}

Having determined the first and second class constraints we can
now proceed to count the number of local degrees of freedom of
this theory. We have $2n$ canonical variables $A_i$ and $2n$
canonical momenta $p^i$. To this number, $4n$, we subtract the
number of second class constraints, namely 2, and twice the
number of first class constraints, $2\times (2n-1)$, so that
\begin{equation}
2 {\cal N } = 4n - 2 -2\times (2n-1) =0.
\end{equation}
Thus, there are no local degrees of freedom. This is a pure
topological field theory and the only degrees of freedom that
may be present are the global ones. We must stress again,
however, that this is a peculiarity of the $N=1$ theory, which
is in that sense a poor representative of the general case.

\subsection{Reduced action}

It is instructive to write down the reduced action once the
second class constraints have been solved. Since the constraints
$\phi^\alpha$ are linear in the momenta, this is easily
achieved.  Upon inserting the solution inside the action, one
obtains a reduced action for the relevant dynamical fields and
Lagrange multipliers,
\begin{equation}
S[A_\alpha, A_p,p^p;A_0, N^p] = 
\int_\Re\int_{\Sigma}[l^\alpha(A_\alpha,A_p) \dot A_\alpha 
+ p^p \dot A_p - A_0  G - N^p \bar H_p ]  ,
\label{action3}
\end{equation}
which must be varied with respect to all its arguments.

The  symplectic structure is not canonical due to the presence
of the factor $l^\alpha (A_\alpha,A_p)$ in the first kinetic
term.  The symplectic form
\begin{equation}
\omega= \frac{1}{2} \int_\Sigma (\Omega^{\alpha\beta}
\mbox{\boldmath $\delta$} A_\alpha \mbox{\tiny $\wedge\,$} 
\mbox{\boldmath $\delta$} A_\beta + 2 \Omega^{\alpha p}
\mbox{\boldmath $\delta$} A_\alpha \mbox{\tiny $\wedge\,$} 
\mbox{\boldmath $\delta$} A_p + 2\mbox{\boldmath $\delta$} 
A_p \mbox{\tiny $\wedge\,$} \mbox{\boldmath $\delta$} p^p)
\end{equation}
is, by construction, invertible and its inverse provides the
Dirac bracket
\begin{equation}
\begin{array}{lll}
\{A_\alpha, A_\beta\}^*=J_{\alpha\beta},
\hspace{7mm}
&
\{A_\alpha, A_p\}^*=0,
\hspace{7mm}
&
\{A_\alpha, p^p\}^*=J_{\alpha\beta}\Omega^{\beta p},
\\
\hspace{1mm}\{A_p, A_q\}^*=0,
\hspace{7mm}
&
\hspace{1.5mm}\{A_p, p^q\}^*=\delta_p^q,
\hspace{7mm}
&
\hspace{1.5mm}\{p^p, p^q\}^*=0,
\end{array}
\end{equation}
which is equivalent to that defined in Eq.
(\ref{Dirac-bracket/1}), in agreement with the general theory
\cite{HT-book}.

\subsection{Comparison with superparticle}

The canonical analysis of the $U(1)$-Chern-Simons theory in
higher dimensions presents many similarities with that of the
superparticle (see, e.g.,  Ref. \cite{Brink} and references
therein).  In both cases, although the original action is
manifestly covariant to begin with, one cannot reach a complete
canonical formulation without breaking explicitly this manifest
covariance.  This is because  one cannot isolate covariantly the
second class constraints \cite{Bengtsson}.  Furthermore,
although one can write down a complete set of first class
constraints that transform covariantly (here, the constraints
$F_{ij}\phi^j$), these first class constraints are redundant,
implying in a BRST treatment the presence of ghosts of ghosts.
Again, one cannot isolate covariantly a complete, {\em
irreducible}, set of first class constraints.

There is another interesting similarity: if one chooses to work
with the covariant, redundant, first class constraints
$F_{ij}\phi^j$, one finds that the reducibility identities are
\begin{equation}
\mu^{ki} F_{ij}\phi^j = 0
\end{equation}
(on the constraint surface $G=0$) with
\begin{equation}
\mu^{ki} = \epsilon^{kii_1\ldots i_{2n-2}}F_{i_1i_2}\ldots 
F_{i_{2n-3}i_{2n-2}}
\end{equation}
These reducibility identities, in turn, are not independent
since $F_{lk} \mu^{ki} = 0$ (weakly), and this reducibility of
the reducibility is itself not irreducible, etc.  One is thus
led to an infinite tower of reducibility identities, requiring
an infinite set of ghosts of ghosts in the BR ST formulation,
exactly as in the superparticle case \cite{Brink,Kall}.

\section{The  $G\times U(1)$ theory}
\setcounter{equation}{0}
\label{G*U(1)}

The Hamiltonian analysis performed so far in the $N>1$ case is
incomplete because $(i)$ the second class constraints have not
yet been eliminated and $(ii)$ in the case of manifolds with
boundaries, it is known that the Hamiltonian has to be
supplemented with some boundary terms \cite{Regge-Teitelboim}, a
problem not yet discussed. These two issues are cumbersome on
the computational side --- even though conceptually easy --- if
one works with an arbitrary Lie group $G$ and, actually, they
cannot be treated in a general covariant way. It is surprising,
therefore, that a drastic simplification takes place if one
couples a $U(1)$ factor to the group $G$. To avoid unessential
technical difficulties, we shall restrict our attention to that
case, which illustrates all the conceptual features.  In this
section, we solve all second class constraints and compute the
Dirac bracket. In the next section, we deal with the boundary
terms necessary to make the Hamiltonian well-defined. For
simplicity, we work explicitly in five dimensions but we shall
indicate how the results obtained here can be extended to any
odd dimensional spacetime.
    
\subsection{The invariant tensor for  $G\times U(1)$ }

Consider the Chern-Simons action in five dimensions for the Lie
group $G\times U(1)$. [In this section capital Latin indices
$A,B\ldots$ run over $G\times U(1)$. Small Latin indices
$a,b\ldots$ run over $G$, and $1$ denotes $U(1)$.] It is
straightforward to see that the invariance condition on the
tensor $g_{ABC}$ implies the following restrictions on its
components. The components $g_{abc}$, $g_{ab1}$ and $g_{a11}$
must separately be invariant under the adjoint action of $G$,
and $g_{111}$ is an arbitrary constant.

We shall now impose three extra conditions on the group $G\times
U(1)$ and its invariant tensors. First we assume that $g_{a11}$
is zero.  Usually, this is  not an additional requirement
because, in general, there is no vector invariant vector under
the adjoint action of $G$.  Second, we assume that $G$ admits an
invariant non-degenerate quadratic form $g_{ab}$, as it is the
case if $G$ is semisimple, and we take
\begin{equation}
g_{ab1} = g_{ab}.
\end{equation}
Finally, we impose
\begin{equation}
g_{111}=0.
\end{equation}
This condition is justified on simplicity grounds since, as we
shall see below, it allows for a simple separation between first
and second class constraints.  Note that the gauge fields
associated with $G$ and $U(1)$ respectively are  not decoupled
in the action, because $g_{ab1} \neq 0$.

\subsection{Dirac Brackets}

An immediate consequence of this choice for the invariant tensor
is that the second class constraints can be explicitly isolated
and solved, at least in a generic region of phase space.

The constraint equations ($K_A=-\frac{1}{2} g_{ABC}
F^B\mbox{\tiny $\wedge\,$} F^C=0$), in this case, are
\begin{eqnarray}
K_a &=& -\frac{1}{2} g_{abc} F^b \mbox{\tiny $\wedge\,$} F^c -
F^1 \mbox{\tiny $\wedge\,$} F_a = 0 \hspace{7mm} (A=a) \\
K_1 &=& -\frac{1}{2} g_{ab} F^a \mbox{\tiny $\wedge\,$} F^b = 0 
\hspace{7mm} (A=1) 
\end{eqnarray}  
An obvious solution for these equations is $F^a=0$ and $F^1$
completely arbitrary. The matrix $\Omega$ evaluated on this
particular solution has the block form
\begin{equation}
\Omega^{ij}_{AB}|_{F^a=0} =  \left(\begin{array}{c|c}   
0   &  0   \\ \hline
0   &  -\frac{1}{2}g_{ab} \epsilon^{ijkl} F^1_{kl}     
\end{array}  \right).
\label{Ome} 
\end{equation}
If the matrix $F^1_{ij}$ is non-degenerate, then $\Omega$
(evaluated on that particular solution) has the maximum rank
$4N-4$. Hence, we have proved that the $G\times U(1)$ theory is
$generic$ in the sense described in Sec. \ref{Generic}. Of
course, a degenerate $F^1$ provides also a solution for the
equations of motion. Such a solution, however, belongs to a
different branch of the theory with a smaller number of local
degrees of freedom.
  
Since the solution  $F^a=0$ and $F^1$ non-degenerate is such
that the matrix $\Omega$ has maximum rank, a sufficiently small
perturbation around it will not change this rank (the rank is a
semi-continuous function  from below). Thus, on the portion of
phase space around the solution $\{ F^a=0, \det{F^1}\neq 0\}$,
the sub-matrix $\Omega^{ij}_{ab}$ is invertible. This means that
among the primary constraints $\phi^i_A$, the subset $\phi^i_a$
is second class.  Moreover, since the matrix $F^1_{ij}$ is
invertible in an open region around the above solution, one can
replace the set of constraints $(\phi_1^i,\phi_a^i)$ by the
equivalent set $(H_i,\phi_a^i)$, where
$H_i=F_{ij}^1\phi_1^j+F_{ij}^a\phi_a^j$ are the first class
constraints that generate the diffeomorphisms. In this new set,
the constraints are separated into first class and second class.
The technical simplification that motivated the choice of group
$G \times U(1)$ appears precisely here:  the separation of the
constraints into first and second classes can be easily
achieved.

Once the second class constraints have been isolated one can
compute the Dirac bracket.  We define the inverse of the matrix
$\Omega^{ij}_{ab}$ by
\begin{equation}
\Omega^{ij}_{ab} J^{ac}_{jk} = \delta^i_k \delta^c_b.
\label{J}
\end{equation}
The Dirac bracket among two phase space functions $A$ and $B$ is
then given by
\begin{equation}
\{A,B\}^* = \{A,B\} - \int_\Sigma dz \{A,\phi^i_a(z) \}\ 
J^{ab}_{ij}(z) \  \{\phi^j_b(z) ,B\}, \label{Dirac} 
\end{equation}  
which gives the following Dirac bracket relations among the
elementary variables:
\begin{equation}
\begin{array}{lll}
\{A^a_i,A^b_j\}^*=J^{ab}_{ij},\hspace{7mm}
& \{A^a_i,A^1_j\}^*=0,\hspace{7mm}
& \{A^a_i,p^j_1\}^*=J^{ab}_{ik}\Omega^{kj}_{b1},
\\
\hspace{1mm} \{A^1_i,p^j_1\}^*=\delta_i^j,\hspace{7mm}
& \{A^1_i,A^1_j\}^*=0,\hspace{7mm}
& \hspace{1mm} \{p_1^i,p^j_1\}^*=0.
\end{array}
\end{equation}
The brackets of the variable $A^1_i$ with the other variables
are simple.  However, the brackets of the variables $A^a_i$
among themselves and with the $p^j_1$'s are more involved.  In
the quantum theory, the $A^a_i$'s are not commuting operators.
An interesting question not investigated here is to find an
explicit realization of the Dirac bracket algebra in terms of
commutators.  This question is not straightforward, because the
left hand sides of the Dirac brackets among the basic variables
are not $c$-numbers.

If we now work with the  Dirac bracket, we can set the second
class constraints strongly equal to zero and keep only the first
class ones in the formalism.  These are given by $G_a,G_1$ and
$H_i$ defined in Eqs. (\ref{G}) and (\ref{Hi}), respectively and
are the generators of internal gauge transformations and of
spatial diffeomorphisms.  After the second class constraints
have been set equal to zero, these generators simplify to
\begin{eqnarray} 
G_a &=& -K_a  \\ G_1 &=&
-K_1 + \partial _i \phi^i_1 \\ H_i &=& F^1_{ij} \phi^i_1 
\label{Hi/G*U} 
\end{eqnarray}
One can easily check that, in the Dirac bracket, $G_a$ and $G_1$
given above satisfy the Lie algebra of $G\times U(1)$ and $H_i$
satisfy the algebra of diffeomorphisms up to a gauge
transformation.

The same analysis can be performed in all odd dimensions greater
than five provided that the invariant tensor $g_{A_1 A_2 \dots
A_{n+1}}$ satisfies the conditions $g_{11 \dots 1}=0$, $g_{a1
\dots 1}=0$, $g_{a b 11\dots 1}= g_{ab}$ (with $g_{ab}$ an
invertible invariant metric), while all the other components are
invariant tensors of $G$. Again, the configuration $F^a=0,
\det{F^1} \neq 0$ is a maximum rank solution for the constraint
equations and therefore the theory is generic.

\section{Global symmetries and WZW$_4$ algebras}
\setcounter{equation}{0}
\label{WZW}

We turn now to the problem of boundary conditions and boundary terms
when the spatial manifold has a boundary, again in the simple context of
a gauge group of the form $G\times U(1)$. There is a special motivation
for doing this. Indeed, a four dimensional analog of the WZW Lagrangian
exists \cite{Nair}. This model is characterized by a symmetry algebra
that generalizes the familiar Kac-Moody algebra in two dimensions, and
which has been called WZW$_4$ algebra \cite{Nair,Moore-}.  It is a
natural question to see whether this algebra  can be obtained from a
Chern-Simons theory in five dimensions just as the Kac-Moody algebra is
generated from the 2+1 Chern-Simons theory \cite{Moore-Sieberg} (see
also Ref. \cite{B}). This problem has been already studied in the
literature in interesting works that considered modifications of the
Chern-Simons action.  In Ref. \cite{Dunne}, the Yang-Mills action was
added to the Chern-Simons action in order to make the symplectic
structure simpler. This procedure, however, breaks the diffeomorphism
invariance.  In Ref. \cite{Nair},  a Chern-Simons theory coupled to a
{\em fixed} K\"alher form is considered. This procedure breaks also part
of the diffeomorphism invariance of the theory. Finally, 4D currents
arising from the Abelian $N=1$ Chern-Simons theory have been
studied in Ref. \cite{Gupta-Stern}. 

It is the purpose of this section to prove that the issue of global
charges in pure Chern-Simons theories can be analysed in the full 
unmodified non-Abelian theory without making any particular assumptions
(other than the boundary conditions). In this paper, however, we do not
study the effective Lagrangians arising at the boundary, but only the
algebras.

As usual in any gauge theory, in the presence of boundaries, the gauge
symmetries split into ``proper" and ``improper" gauge symmetries. The
proper symmetries are those transformations which are generated by the
constraints through Poisson brackets. Their generators are thus weakly
zero. Improper symmetries, on the other hand, are generated by the
constraints supplemented with a (non-vanishing) boundary term. They
should be viewed as global symmetries. After the gauge is fixed and the
constraints are strongly set equal to zero, these boundary terms, the
charges, `survive' and satisfy a well-defined Dirac bracket algebra at
the boundary.

Consider the Chern-Simons theory for the group $G\times U(1)$
discussed in the last section and consider a gauge
transformation along $G$ with a parameter $\eta^a$.  We define
\begin{equation}
G_Q(\eta) = \int_\Sigma \eta^a G_a + Q(\eta)
\label{K(Q)}
\end{equation}   
where the charge $Q$ has to be adjusted so that $G_Q(\eta)$
generates the transformation $\delta A^a =- D\eta^a$ for the
gauge field, even at the boundary. By direct application of the
Dirac bracket defined in Eq. (\ref{Dirac}), the transformation
induced by $G_Q(\eta)$ on $A^a_i$ is given by,
\begin{equation}
\delta_\eta A^a_i(x) = \{A^a_i(x), G_Q(\eta) \}^* 
= J^{ab}_{ij}(x) \frac{\delta G_Q(\eta)}{\delta A^b_j(x)}.
\label{delta-A}
\end{equation}
Recall that since we are working with the Dirac brackets, the
generator $G_a$ (for the group $G\times U(1)$) has the simple
expression $G_a = -K_a$.

We need to compute the functional derivative of $G_Q(\eta)$. From 
the definition of $G_a$, one finds that if the charge $Q$
satisfies the equation
\begin{equation}
\delta Q = \int_{\partial \Sigma} \eta^a (g_{abc} 
F^a\mbox{\tiny $\wedge\,$} \delta A^b 
+ g_{ab}  F^1 \mbox{\tiny $\wedge\,$} \delta A^b),
\label{delta-Q}
\end{equation}
then, the derivative of $G_Q(\eta)$ is well-defined and given by
\begin{equation}
\frac{\delta G_Q(\eta)}{\delta A^a_i(x)} = 
-\Omega^{ij}_{ab}(x)  D_j \eta^b(x).
\label{d-K/d-A}
\end{equation}
Formula (\ref{delta-A}), together with  Eq. (\ref{J}) gives
$\delta A^a_i = -D_i \eta^a$, as expected.

There are two remaining things to be checked before we can fully
promote $G_Q(\eta)$ to  the generator of gauge transformation
with a parameter $\eta$.  First, we need to integrate relation
(\ref{delta-Q}) in order to extract from it the value of $Q$.
Second, we need to compute the algebra of $G_Q(\eta)$.

In order to integrate Eq. (\ref{delta-Q}), we shall impose the
following boundary conditions,
\begin{equation}
F^a=0, \hspace{7mm}  F^1 = \omega \equiv \hbox{ fixed
2-form},  \hspace{7mm} (\mbox{at  the boundary}).
\label{boundary-cond}
\end{equation}
The charge $Q$ for these boundary conditions is then given by
\begin{equation}
Q(\eta) = \int_{\partial \Sigma} \omega \mbox{\tiny $\wedge\,$} 
A^a \ \eta^b g_{ab}.
\label{Q/G*U}
\end{equation}  

Now we turn to the problem of the algebra of $G_Q(\eta)$.  The
Dirac bracket of two generators $G_Q(\eta)$ and $G_Q(\rho)$ is
\begin{equation}
\{ G_Q(\eta), G_Q(\rho) \}^* = 
\int_\Sigma \Omega_{ab} \mbox{\tiny $\wedge\,$} D\eta^a 
\mbox{\tiny $\wedge\,$} D\rho^b.
\label{K-K}
\end{equation}
After an integration by parts, keeping all boundary terms, and
using the Jacobi identity for the structure constants $f^a_{\
bc}$ of the group $G$, the right hand side of Eq. (\ref{K-K})
can be written as
\begin{equation}
\{ G_Q(\eta), G_Q(\rho) \}^* = \int_\Sigma [\eta,\rho]^a G_a 
+ \int_{\partial \Sigma} \omega \mbox{\tiny $\wedge\,$} 
[\eta,\rho]_a A^a + \int_{\partial \Sigma} \omega 
\mbox{\tiny $\wedge\,$} \eta_a d\rho^a 
\label{K-K2}
\end{equation}
where $[\eta,\rho]^a = f^a_{\ bc} \eta^b \rho^c$.  The boundary
term in the right hand side has two pieces.  The first term is
precisely the charge $Q([\eta,\rho])$ that regularizes the bulk
integral. The second term, on the other hand, does not depend on
the fields that are varied at the boundary and, therefore, is a
central term.  The algebra (\ref{K-K2}) can then be rewritten in
its final form,
\begin{equation}
\{ G_Q(\eta), G_Q(\rho) \}^* = G_Q([\eta,\rho]) + 
\int_{\partial \Sigma} \omega \mbox{\tiny $\wedge\,$}  
\eta_a d\rho^a
\label{K-K3}
\end{equation}
As we can see, this algebra  {\em is not} homomorphic to the
original algebra of $G$ but it is a non-trivial  central
extension of it.  (The possibility of non-trivial central
charges in the canonical realization of global charges given by
surface integrals was demonstrated in general in Ref.
\cite{Brown-Henneaux1}.)  This algebra was first obtained in Ref.
\cite{Nair}.  In that paper, the role of
$F^1 \equiv \omega$ was played by a non-dynamical K\"ahler form
while here, it appears as the curvature of the $U(1)$ factor.

The algebra (\ref{K-K3}) is a natural generalization of the
Kac-Moody algebra existing in two dimensions. We have shown in
this section that, as one could have expected, Chern-Simons
theory in five dimensions generates a `conformal' theory on the
four dimensional boundary.

This analysis can be repeated in higher dimensions with an
invariant tensor that fulfills the conditions spelled out in the
previous section.  One finds that  the algebra of the charges is
simply
\begin{equation}
\{ G_Q(\eta), G_Q(\rho) \}^* = G_Q([\eta,\rho]) + \int_{\partial \Sigma}
\omega \mbox{\tiny $\wedge\,$} \omega \mbox{\tiny $\wedge\,$}
\dots  \mbox{\tiny $\wedge\,$} \omega \mbox{\tiny $\wedge\,$}  \eta_a d\rho^a,
\label{2ndim}
\end{equation} 
where $\omega$ is a fixed two-form.  In analogy with the
four-dimensional terminology, it may be  called ``WZW$_{2n}$
algebra".

\section{Conclusions}
\setcounter{equation}{0}
\label{conclusions}

We have shown that higher dimensional Chern-Simons theories,
even though constructed along the same topological pattern as in
$2+1$ dimensions, have local degrees of freedom provided that
the invariant tensor that enters the action fulfills an
appropriate generic condition.  This condition implies that
there are no accidental gauge symmetries, so that the number of
gauge symmetries grows more slowly  with the dimension of the
gauge group than with the number of dynamical variables.  This
result cannot be  anticipated by analysing the case of a single
abelian field, which is not representative of the general case.

Chern-Simons theories in higher dimensions provide, accordingly,
examples of theories that are generally covariant without
involving a dynamical metric, and yet, that carry local
dynamical degrees of freedom.  Therefore, they constitute
counter-examples to the belief that such theories can contain
only global or surface degrees of freedom.

We have illustrated the presence of local degrees of freedom
with examples in seven dimensions.  These examples complement
the five dimensional examples given in Ref. \cite{BGH}.  We have
also applied the analysis to Lovelock-Chern-Simons gravity in
any odd dimensions and have established by a mere count of the
number of local degrees of freedom that the first order
(Palatini) and the second order (metric) formalisms are not
equivalent.

We have also shown that the timelike diffeomorphisms do not lead
to independent constraints.  The implications of this remarkable
feature for the quantum theory remain to be explored.  As a
first step, it would be interesting to investigate how the loop
representation must be defined when the connection obeys the non
trivial Dirac brackets computed above.

We have finally studied the global charges that naturally arise
in the presence of boundaries and have shown that, at least for
the gauge group $G\times U(1)$, the gauge generators satisfy the
WZW$_4$ algebra  in five dimensions, just as the Kac-Moody
algebra arises in manifolds with boundaries for Chern-Simons
theories in 2+1 dimensions. This WZW$_4$ algebra can be
generalized to the WZW$_{2n}$ algebras, which appear again as
global symmetry algebras on the boundary for Chern-Simons
theories in dimension $2n+1$.

\section*{Acknowledgments}

M.H. is grateful to LPTHE (Universit\'es Paris VI and Paris VII)
for kind hospitality.  M.B. is partially supported by a grant from 
Fundaci\'on Andes (Chile), grants 1930910-93 and 1960065-94 from 
FONDECYT (Chile), and by institutional support to the
Centro de Estudios Cient\'{\i}ficos de Santiago provided by
SAREC (Sweden) and a group of Chilean private companies
(EMPRESAS CMPC, CGE, COPEC, MINERA LA ESCONDIDA, NOVAGAS
Transportandores de Chile, ENERSIS, BUSINESS DESIGN ASS., XEROX
Chile).  L.J.G. was supported by funds provided by DGICYT and
MEC (Spain) under Contract Adjunct to the Project No.
PB94--0107.  The work of M.H. is partially supported by research
funds from F.N.R.S. (Belgium) and a research contract with the
Commission of the European Community.

\end{document}